\documentclass{amsart} 
\usepackage{mathrsfs,amssymb,amsmath,amsfonts,amsxtra}
\usepackage{graphicx,color,verbatim}
\usepackage[colorlinks]{hyperref}

\begin{document}
\sloppypar \sloppy
\title[An anwer to the Wallstrom objection]{An answer to the Wallstrom objection against Nelsonian stochastics}
\author{I. Schmelzer}
\thanks{ilja.schmelzer@gmail.com\\Akeleiweg 7\\D 12487 Berlin, Germany}

\begin{abstract}
A serious objection made by Wallstrom against quantum interpretations based flow variables, in particular Nelsonian stochastics, is their empirical inequivalence with quantum theory: They are unable to obtain a quantization condition for flows around zeros which is automatically fulfilled by the wave function. 

It is found that the quantization condition follows from a simple additional postulate: The Laplace operator of the density has to be finite and positive at zeros of the density. This postulate is a quite natural consequence of subquantum theories, as far as they conform to a simple principle of minimal distortion of the quantum solutions.  
\end{abstract}

\maketitle


\newcommand{\pd}{\partial} 
\newcommand{\ud}{\mathrm{d}} 
\newcommand{\f}{\varphi}
\newcommand{\w}{\nabla\times v}
\renewcommand{\a}{\alpha}

\renewcommand{\H}{\mbox{$\mathcal{H}$}} 
\newcommand{\B}{\mbox{$\mathbb{Z}_2$}} 
\newcommand{\Z}{\mbox{$\mathbb{Z}$}}
\newcommand{\R}{\mbox{$\mathbb{R}$}}
\newcommand{\C}{\mbox{$\mathbb{C}$}}

\newtheorem{theorem}{Theorem}
\newtheorem{principle}{Principle}
\newtheorem{postulate}{Postulate}
\newcommand{\Sch}{Schr\"{o}dinger\/ }

\section{Introduction}

\subsection{Quantum interpretations based on flow variables} 

There is a remarkable class of interpretations of quantum theory based on variables often named ``hydrodynamic'', which I prefer to name ``flow variables''\footnote{To name them ``hydrodynamic'' is misleading, because a hydrodynamic interpretation is possible only in the uninteresting one particle case. An interpretation in terms of a probability flow is much more natural. To name them ``flow variables'', and the interpretations based on them ``flow interpretations'', seems therefore more appropriate. It leaves the question of the nature of the flow -- probability flow or whatever -- undecided. What characterizes these interpretations is that the flow variables $\rho(q), v^i(q)$ are considered to be more fundamental than the wave function $\psi(q)$, not the particular interpretation given to the flow variables.} -- a density $\rho(q)$ and a velocity field $v^i(q)$. The velocity field has a potential
\begin{equation}\label{guiding}
v^i(q)=\delta^{ij}\pd_j S(q),
\end{equation}
and this potential $S(q)$, together with the density $\rho(q)$, define the wave function by the polar decomposition
\begin{equation}\label{polar}
\psi(q)=\sqrt{\rho(q)}e^{\frac{i}{\hbar}S(q)}.
\end{equation}
The use of these variables goes back to the time of invention of quantum theory, in particular to Madelung \cite{Madelung}, de Broglie \cite{deBroglie}, and Bohm \cite{Bohm}. It is also the base of Nelsonian stochastics \cite{Nelson}, where $v^i(q)$ becomes an average velocity.

\subsection{The Wallstrom objection}

The objection made by Wallstrom \cite{Wallstrom} is formally simply a result about empirical inequivalence between the \Sch equation and the corresponding equations in terms of $\rho(q)$ and $S(q)$. While the continuity equation 
\begin{equation}\label{continuity}
\partial_t \rho + \pd_i(\rho v^i) = 0.
\end{equation}  
is unproblematic, the equation for $S(q)$ 
\begin{equation}\label{Bohm}
-\pd_t S(q) = \frac12 \delta^{ij}\pd_i S(q)\pd_j S(q) + V(q) -\frac{1}{2} \frac{\Delta \sqrt{\rho}}{\sqrt{\rho}}.
\end{equation}
becomes meaningless at $\rho(q)=0$. This would be harmless in itself, but in quantum theory for closed paths around the zeros the following ``quantization condition'' holds:
\begin{equation}\label{curlquantization}
\oint \delta_{ij} v^i(q) dq^j = \oint \pd_j S(q) dq^j = 2\pi m\hbar, \qquad m\in\Z.
\end{equation}
In terms of the wave function defined by \eqref{polar}, this condition is a triviality: It has to be a uniquely defined, continuous function. But formulated in terms of the flow variables, it becomes completely unmotivated and artificial, and does not follow from the equations. To derive it from a subquantum theory, a theory where the wave function does not even exists, seems hopeless. As Wallstrom \cite{Wallstrom} writes, to ``the best of my knowledge, this condition [\eqref{curlquantization}] has not yet found any convincing explanation outside the context of the \Sch equation''. 

The aim of this paper is to give another answer to the Wallstrom objection. It is shown that the quantization condition \eqref{curlquantization} can be obtained from a subquantum theory in flow variables. In particular, the following regularity postulate
\begin{postulate}[regularity of $\Delta\rho$ at zero]\label{p1}
If $\rho(q)=0$, then $0 < \Delta \rho(q) < \infty$ almost everywhere,
\end{postulate}
which excludes only the two extremal values $0$ and $\infty$ as irregular, is already sufficient to enforce the quantization condition. And it is shown that this conditions can be expected to be derived from a subquantum theory. It follows even almost automatically. I propose a ``principle of minimal distortion'' of the quantum solutions which, I argue, plays a role analogical to the null hypothesis in statistics: Properties of the quantum solution which do not have to be changed to get rid of infinities will not be changed. And this variant of the null hypothesis is already sufficient to derive the postulate. 

\section{The relevance of the Wallstrom objection}

But, first, the relevance of this objection has to be discussed, and where a previous attempt by Smolin \cite{Smolin} to answer it has failed. 

\subsection{The ignorance answer} 

The reason is that one possible answer to the Wallstrom objection is simply ignorance: One accepts the mathematics, and simply adds the quantization condition \eqref{curlquantization} to the list of postulates, but otherwise ignores it as ``something of a non-problem''. The condition is unjustified, but so what? As an anonymous referee has explained it: ``In this regard the hydrodynamic approach is in no worse situation than the usual quantum formalism which likewise is unjustified (why should a probability amplitude be single-valued in Euclidean space? Why should it obey the Schrodinger equation?)''.

Ignorance is even more justified if one is not even interested in interpretational questions, but simply uses the flow variables as another choice of variables, a choice which may be, for one reason or another, more appropriate for a particular problem. And so it is quite natural that, as the referee has observed, ``Wallstrom's paper has been largely ignored by the community who use the fluid theory and the approach is very widely employed, from superfluidity to chemical physics.''

But there is a domain of research where the Wallstrom objection is important. It is not the domain of practical application of the flow variables. And it is even not the whole domain of use of the flow variables in the interpretation of quantum theory. It is only some subset of interpretations based on flow variables where the Wallstrom objection is relevant. But it is, arguably, the most interesting and physically most important subset of such interpretations. 

\subsection{About the physical relevance of interpretations}

Now, many physicists think that interpretations are only metaphysics, and therefore not physically important at all, so the very phrase ``physiclly most important'' is already meaningless. But they are wrong: At least some interpretations are physically important. Even if they do not make empirical predictions different from those of the theory which is interpreted, they may define research programs for the development of different, more fundamental theories. 

Roughly speaking, we can distinguish two types of interpretations. The first type could be named ``orthodox'': The mathematics of the theory and the empirical predictions are not questioned at all, they are considered by these interpretations (but not necessarily by their proponents) as fundamental, as absolute truths. Different orthodox interpretations may differ about the particular variables they accept as beables and about their nature, but they don't give any hints for a development of different, more fundamental theories. 

The other type of interpretation can be named ``heretical''. They are also interpretations, thus, they do not propose different equations and do not make different empirical predictions. But it is the very interpretation which tells us that the equations of the theory are not exact, and that the empirical predictions of the theory have to fail somewhere. So it follows from these interpretations that there has to be some different, more fundamental theory. 

Even more, these interpretations usually specify some properties of the more fundamental theory. For example, they can specify the ontology of the more fundamental theory, even some of the equations (those of the original theory which can be left unchanged), or specify that some predictions of the interpreted theory have to fail somewhere. Thus, they define a research program: A program for theoretical physics -- the development of the more fundamental theory which has to replace the original theory. And possibly also a program for experimental physics -- to focus the interest on those particular effects where, according to the interpretation, the theory becomes wrong. 

Such research programs are clearly physically important. Even if one insists that interpretations themself are only metaphysics -- the heretical interpretations are, in this case, examples of physically important metaphysics. 

\subsection{Flow interpretations as research programs for a subquantum theory} 

Now, the flow interpretations are, in a quite natural way, interpretations of the second, heretical type. The decisive point is the potentiality condition \eqref{guiding}. 

The potentiality condition is known from standard hydrodynamics as an approximation for low energies. So the very association with hydrodynamical variables already suggests that the potentiality condition is not fundmental. 

But the decisive point is not that analogy, which is anyway misleading in some points, but the very fact that the potentiality condition is violated at the zeros of the quantum wave function already in pure quantum theory.  The zeros move in time, and there may be more or less of them, so these violations of the potentiality condition which are already present in quantum theory cannot be described in a simple way as caused by some external circumstances. 

And, even worse, the curl as well as the velocities themself become infinite near the zeros. Infinities in physically relevant variables are always a strong hint for modification, so the very property that the velocity $v^i(q)$ is interpreted as relevant strongly suggests -- of course only to those who follow this interpretation -- that quantum theory is only an approximation of some subquantum theory, and that this subquantum theory regularizes the infinities of $v^i(q)$ as well as of the curl. 

The very fact that the infinities of the velocities and the curl correspond to points of zero density have, by the way, their analogy in hydrodynamics:  A hurrican gives a quite similar picture. Of course, only approximately, for distances large enough in comparison with the eye of the hurrican, But the center of the hurrican is also the region of lowest pressure. 

So it is quite natural to classify flow interpretations as being of the heretical type: They define a research program for a subquantum theory. In this subquantum theory, the potentiality condition \eqref{guiding} no longer holds, and, therefore, no wave function exists. Instead, the flow variables $\rho(q)$, $v^i(q)$ exist and define some probability flow, which fulfills the continuity equation \eqref{continuity}.

And the flow interpretations even suggest a domain where quantum theory fails -- the region near the zeros of the wave function, where the $v^i(q)$ become infinite, but will have to be regularized by the subquantum theory, in a way similar to the eye of a hurrican. 

\subsection{The relevance of the Wallstrom objection}

But for this quite natural, specific and interesting research program for a subquantum theory the Wallstrom objection is a really serious problem.  

In fact, if in the subquantum theory the flow is not a potential one, the integrals of type \eqref{curlquantization} over closed loops do have non-integer values. Indeed, they have small values for small circles around regions without zeros of the density inside, with zero limit if the radius of the cicle goes to zero, but with some non-zero (and therefore non-integer) values, because, else, the potentiality condition would hold. So the quantization condition \eqref{curlquantization} cannot 
hold in the subquantum theory, it has to be derived. 

As a consequence, interesting proposals for subquantum theories have to be rejected as defective. For example, Valentini \cite{Valentini} writes: ``\ldots the deterministic models of refs. [here \cite{Smolin1,Smolin2,Smolin3}] seem to yield derivations of Nelsonian mechanics, but not of quantum mechanics. Some basic element is missing. One must somehow ensure that the circulation of $\nabla_i S$ around nodes of $\rho$ can be non-zero but always restricted to integer multiples of $2\pi$. (And if one wishes to derive the wave function, then of course one cannot simply assume at the outset that $S$ is the phase of a complex-valued field.)''

Another consequence is that variants of the interpretations based on flow variables are proposed which no longer claim to derive the \Sch equation but presuppose it. For example, Bacciagaluppi \cite{Bacciagaluppi} explicitly justifies such a modification of Nelsonian stochastics \cite{Nelson} and Davidson's generalization of it \cite{Davidson}. As well, almost all modern presentations of de Broglie-Bohm theory don't use flow variables.\footnote{While there is the independent argument that this simplifies the introduction into dBB theory for those who already know quantum theory, Pearle and Valentini \cite{Pearle} name the Wallstrom objection a ``decisive'' one against the ``hydrodynamical'' interpretation and mention explicitly that a version of dBB theory which regard the wave function as a basic entity does not have this problem.} 

\subsection{Smolin's proposal} 

One attempt to answer the Wallstrom objection has been made by Smolin \cite{Smolin}. Because only the simplest case of a single degree of freedom on a circle $S^1$ is considered, it has been criticized as not sufficiently general already by Valentini \cite{Valentini}. But it should be rejected as wrong even for this simple case. 

Smolin's proposal is to introduce discontinuous wave functions into the consideration. He notes that they are valid elements of the Hilbert space $\mathcal{L}^2(S^1)$, and so they can be used as initial values for the \Sch equation too. 

But these solutions are simply not equivalent. The wave function $e^{i\a\f}$, $0\le\f<2\pi$, for non-integer $\a$ defines a wave function with discontinuity at $\f=0$. Smolin acknowledges that this wave function is not an eigenstate of the momentum. So, for the simplest case of $H=p^2$, it is also not an eigenstate of energy, and, therefore, the resulting solution of the \Sch equation is not time-independent. In particular, it will be quite singular near $t=\f=0$, but it seems reasonable to expect that the solution will be smooth everywhere else. The corresponding solutions in the flow variables exist too, and will have similar properties -- singular near $t=\f=0$, and time-dependent.

But in the flow variables there is also a time-independent solution of the equations -- namely $\rho(\f,t)=\frac{1}{2\pi}$, $v(\f,t)=\a$. It is this time-independent solution which is not represented by any wave function, and discontinuous wave functions are of no help here. The empirical inequivalence remains. 

\section{The proposed solution}

Let's consider now the question how the regularity postulate \ref{p1} allows to solve the problem. 

\subsection{The restriction to simple zeros}

It should be noted that not all solutions of the \Sch equation fulfill the regularity postulate. Multiple zeros of the wave function, like $\psi(z)=z^2$, give $\Delta\rho = 0$, and are thereforre excluded by the postulate. As a consequence, the integral over a closed path around a single zero gives only values $m=\pm 1$.

But this does not endanger empirical viability of a subquantum theory which allows to derive this regularity postulate in its quantum limit. First, observe that integrals over closed paths, without the additional specification that there is only one zero inside, can give arbitrary integer values $m$. All one needs is just an appropriate number of zeros, with appropriate signs, inside. An the additional restriction that there is only a single zero inside cannot be empirically verified -- there may be, instead, several zeros sufficiently close to each other to be empirically indistinguishable.

Here it is useful to remember that we are interested in interpretations of the second, heretical type, which consider quantum theory as an approximation derived from some subquantum theory. As a consequence, quantum theory has to be recovered only approximately. For interpretations of the orthodox type, which consider quantum theory as exactly true, the situation would be different -- the regularity postulate would exclude some exact, true solutions of quantum theory. But this is not the type of interpretation considered here -- ignorance is already sufficient for them. 

The point that for every physically relevant solution of the \Sch equation there exists, in an arbitrary small environment, another solution which fulfills the regularity postulate can be made, if necessary, mathematically exact. Energy eigenstates are smooth, those with too large energy can be excluded as physically irrelevant, leaving only some extremely large but finite number of them. Then, all what is relevant are linear combinations of them, which are, therefore, smooth too. And for smooth functions, we can apply the apparatus of differential geometry named ``general position''. This tells us that every smooth wave function contains, in an arbitrary small environment, wave functions such that the zeros define a submanifold of codimension $2$, and multiple zeros appear only as transversal self-intersections of this submanifold, and have, therefore, codimension $4$ in the whole configuration space, and codimension $2$ in the submanifold of the zeros. So almost all zeros are simple zeros, that means, the regularity postulate holds. 

\subsection{How the regularity postulate solves the problem}

Let's consider at first the simplest case -- rotationally invariant, stable solutions on the plane $z = x+iy = re^{i\f}$, for the \Sch operator $\hat{H} = -\Delta$. 

For stability in time, there can be no radial velocity component, thus, no dependence of $S(z)$ on $r$, and rotational invariance requires $S(z)=\a\f$ for a constant $\a$. The density cannot depend on the angle, so we have $\rho(z)=\rho(r)$.  The corresponding wave function 
\begin{equation}
\psi(z) = \sqrt{\rho(r)}e^{i \alpha\f}.
\end{equation}
is well-defined on the whole plane only for integer $\a$. Nonetheless, the equations for the flow variables make sense also for general $\a$, and have the solutions 
\begin{equation}
\rho(z) = r^{2|\alpha|}. 
\end{equation}
This corresponds to a wave function $\psi(z) = (x\pm iy)^{|\alpha|}$, which is, indeed, not uniquely defined on the whole plane except for integer $\a$. The solutions we observe are only those described by quantum theory, thus, those with integer $\a$. 

So the problem for the flow interpretations is to explain why $\a$ has to be integer, that means, to exclude all non-integer values of $\a$, despite the very fact that we have nice solutions $\rho(z)=r^{2\a},S(z)=\a\f$, which fulfill all the equations everywhere except the very point where $\rho(z)=0$.  

But, as promised, the regularity postulate \ref{p1} restricts the solutions exactly to those with $|\a|=1$.\footnote{The value $\a=0$ remains also legitimate, but is simply not a zero.} The condition $\Delta \rho(0)<\infty$ excludes $|\alpha|<1$, and the condition $\Delta \rho(0)>0$ excludes correspondingly $|\alpha|>1$. 

\subsection{More general situations} 

The problem is a purely local one -- only a small environment of the zeros matters. In fact, the problem is defined by the parameter $\a$, which is defined by the integral $\a = \frac{1}{2\pi}\oint \nabla S(z) dz$ for a simple closed path around the zero, and which is independent of the particular choice of the path.  

This suggests a technique how to approach the general case: First, one uses a minor modification to obtain general position. In general position, the zeros of the wave function in an $n$--dimensional configuration space will be localized on an ($n-2$)--dimensional submanifold. Self-intersections, corresponding to higher order zeros, appear only on a submanifold of dimension $n-4$, which can be ignored given the ``almost everywhere'' clause (introduced especially for them into the regularity postulate). Then, appropriate local coordinates $(x,y,\dots)$ may be introduced such that the first two coordinates $x=y=0$ define the zero submanifold.  

Then, once $\a$ is known for a (possibly multiply-defined) solution $\psi(q)$, one can define the function $\tilde{\psi}(q)=(x\pm iy)^{-|\a|}\psi(q)$. The $\tilde{\a}$ corresponding to $\tilde{\psi}$ is already zero. As a consequence, $\tilde{\psi}$ is already uniquely defined. It corresponds to the ansatz $S(q)=\a\f+\tilde{S}(q)$, $\rho(q)=r^{2|\a|}\tilde{\rho}$, with $\tilde{S}(q)$ being continuous at zero. 

Let's prove now that $\tilde{\rho}(0)>0$. We put this ansatz into the equation \eqref{Bohm}. It splits there into several parts: A part which gives the same equation for $\tilde{S},\tilde{\rho}$, which is non-singular at least in $\tilde{S}$, a part which contains derivatives only of $\a\f,r^{2|\a|}$, which gives zero as a solution for $V=0$, and an additional mixed part which is singular: 
\begin{equation}
\alpha r^{-1} \pd_\f\tilde{S} - \frac12 \alpha r^{-1} \pd_r \ln \tilde{\rho},
\end{equation}
with $\tilde{S}$ being regular at $r=0$. To compensate this $r^{-1}$ singularity in the limit $r\to 0$, $\ln \tilde{\rho}$ has to be regular at zero too, thus, it follows that $\tilde{\rho}(0)>0$. 

One could think that a singularity could be compensated by a singular quantum potential term for $\tilde{\rho}$ itself. But such a compensation is possible only if $\tilde{\rho}(r)\to\infty$ for $r\to 0$. In particular an $r^{2\beta}$ ansatz gives, for the resulting $r^{-2}$ singularity, $\beta(\beta+2|\a|)=0$, where only $\beta=0$ gives a bounded (but nonzero) density $\rho$.

As a consequence, the solution in the general case has the same $\rho(q)\sim r^{2|\a|}$ behaviour at zeros as in our simple case. And, therefore, the regularity postulate gives the same $|\a|=1$ condition. 

Thus, the general case does not give anything different. So, in the remaining part of the paper we can restrict ourself to the simple case of a rotationally invariant, stable, and two-dimensional problem with zero potential. 

\section{The justification of the postulate}

Of course, one can argue that the regularity postulate has been made up to do the job. If the dependence would have been, instead, $\rho(q)=r^{4\a}$, I could have proposed a criterion using, say, $\Delta^2\rho$. And, in fact, I have to admit that I have made it up that way. But it doesn't matter how the postulate has been guessed. The problem is a different one: Does a subquantum theory based on flow variables have a reasonable chance to derive the regularity postulate or not?  

Moreover, made up or not, we have already reached a certain progress: It is one thing to restrict a value which, conceptually, seems completely arbitrary, like $\oint v^i dq^i$, to integer values, and a quite different one to exclude only the two extremal values $0, \infty$ of an expression like $\Delta\rho$, which even appears in the equations. In my opinion, if the first one is a problem of a ``oh, that's hopeless'' type, the second one is more close to ``and what's the problem with this''. 

Nonetheless, in the remaining part of the paper I will try to give some better justification for the regularity postulate. 

\subsection{The role of $\Delta\rho$ in the energy balance}

First, given that $\Delta\rho$ plays a central role in our considerations, let's give it a more prominent role in the equations. So we can rewrite the quantum potential
\begin{equation}\label{Qdef}
Q[\rho] = -\frac12 \frac{\Delta\sqrt{\rho}}{\sqrt{\rho}} = -\frac14\frac{\Delta\rho}{\rho} + \frac12 u^2,
\end{equation}
where $u$ is the osmotic velocity
\begin{equation}
u^i = \frac12 \pd_i\ln \rho.
\end{equation}
Then, let's modify equation \eqref{Bohm}. Once we consider stable solutions and zero potential, we can omit the time derivative and $V(q)$. We replace $\pd_i S(q)$ by the velocities $v^i(q)$. Then we multiply the whole equation with $\rho(q)$. This gives a sort of balance equation for energy densities:
\begin{equation}\label{balance}
\frac12 \rho v^2 +  \frac12 \rho u^2 = \frac14\Delta\rho.
\end{equation}
The terms on the left hand side are energy densities, and, therefore, the term on the right hand side plays a role in an energy balance, and can be characterized as a sort of energy density itself. 

This is already another large step toward a solution of the Wallstrom objection: The term we have to postulate as regular is not a quite arbitrary mathematical expression, but a term which appears as an energy density in an energy balance equation. Of course, densities in principle may have infinities, but I think there is not much reason to doubt that a subquantum theory \emph{can} justify that all energy densities which appear in it have to be finite everywhere.

Moreover, it is worth to be noted that $\Delta \rho$ is the only term on the right hand side, while all terms on the left hand side are non-negative. This does not give, yet, anything decisive -- we know anyway that $\Delta\rho$ is non-negative at $\rho=0$, and the two terms on the left hand side may become zero at the origin (and, in fact, they become zero, as shown below). Nonetheless, all the subquantum theory has to do to justify $\Delta\rho>0$ is to add some positive term on the left hand side.

So already the quantum equations themself suggest that the remaining problem is solvable one.   

\subsection{The principle of minimal distortion} 

Before speculating about subquantum theory, let's consider the methodological basis for such speculations. Speculations should not be completely arbitrary. There are, in fact, rules of plausible reasoning (see, for example, Jaynes \cite{Jaynes}). In particular, in statistics there is a preference for the null hypothesis: If we have no information which suggests a correlation of two random variables, we prefer the hypothesis of independence: The probabilities of one variable do not depend on the values of the other one. 

An analogical rule in our case would be that, as long as there is no information suggesting something different, qualitative properties of the solution do not change if we switch from quantum to subquantum theory. 

The information suggesting that something has to be changed by subquantum theory is given by the singularity of the equation \eqref{Bohm} near the zeros of the density: The velocity $v(q)$ becomes infinite at $\rho(q)=0$. Here one can identify another general principle, applicable to many cases of more fundamental theories: They lead to essential modifications where the less fundamental theory fails, but these modifications become less and less important in regions far away from the failures. 

Above principles share something one can name ``minimal distortion'' -- one should not distort the solutions of the less fundamental theory without necessity. So let's formulate them in form of the following

\begin{principle}[minimal distortion] The regularization will get rid of all infinities, but change only what is necessary for this purpose. Any qualitative properties which do not have to be changed remain unchanged. Moreover, the distortion by regularization will be larger near the infinities.
\end{principle} 

Such a general principle is extremely helpful for speculations, because it makes a difference between properties which can be assumed without further justification and those which require further justification. Assuming the null hypothesis in statistics does not require any justification, assuming a correlation requires justification. And, similarly, what follows from the principle of minimal distortion proposed here does not require further justification, while any claim violating this principle would require justification. 

Or at least this is what I think, and what has motivated me to formulate and justify this methodological principle separately. A simple ``let's assume that \ldots'' presentation would not have allowed me to make this point.

\subsection{How the regularized solution probably looks like}

So let's look now at the regularization of the solutions to be provided by subquantum theory. The velocity $v$ should not only be finite at $r=0$, it should be zero, for obvious symmetry reasons. The singularity of $v$ is $|v|= \a r^{-1}$, so to regularize down to zero it we have to multiply it with a function of sufficiently high power of $r$. For example, the regularization 
\begin{equation}\label{velocityregularization}
v \to \tilde{v}=v\frac{r^2}{r_0^2+r^2}
\end{equation}
would do the job. As required, $|v(0)|=0$, and the distortion of $v$ decreases for $r\to\infty$.  

The next interesting property is the curl $\w$ of the solution. For the quantum solution, it was infinite in the origin but zero everywhere else. With \eqref{velocityregularization}, this automatically changes. But, in agreement with our principle of minimal distortion, we can assume that $|\w|$ remains small for large $r$ and that it has its maximum at $r=0$. This is in fact what happens for the particular regularization \eqref{velocityregularization}:  We obtain
\begin{equation}
\w = \alpha \frac{2r^2_0 }{(r^2+r^2_0)^2}
\end{equation}

Next, the osmotic velocity $u=\frac12\nabla\ln \rho$ is singular too. The reason for this singularity is that the density at zero is zero. Some small but non-zero density would correct this. But there is also a physical reason to assume that the density becomes non-zero: It is the heavy rotation around the center which causes the small density. But once the velocity in the center becomes zero and $\w$ finite, this will be no longer sufficient to enforce a zero density at the center. So, given that the curl is maximal there, we can reasonably expect a minimal density at the center, but not zero density.

So there will be also a regularization of the density 
\begin{equation}
\rho\to \tilde{\rho} = \rho + \delta\rho.
\end{equation}
The mathematically most trivial regularization procedure would be a simple shift by a constant 
\begin{equation}
\rho\to\tilde{\rho}=\rho+\rho_0.
\end{equation}
Remarkably, it leaves $\Delta\rho$ completely unchanged. 

But the regularized solution will be a solution of the fundamental subquantum equations, not the result of the simplest imaginable regularization procedure. So we should not expect that $\delta\rho=\rho_0$ is constant. Following the principle of minimal distortion we have, instead, to assume that $\delta\rho$ is maximal at the center and decreases with $r$. This has an interesting consequence: $\Delta\delta\rho\le 0$, so that $\Delta \rho \ge \Delta\tilde{\rho}$. On the other hand, following again the principle of minimal distortion, the regularization will leave the minimum of the density at $r=0$. Thus, despite the fact that $\tilde{\rho}(0)>0$, and, correspondingly, $\Delta\tilde{\rho}<0$ would be possible in principle, we can assume that it remains a minimum, so that $\Delta\tilde{\rho}\ge 0$. 

So the minimal distortion principle, applied to the regularization of density, is already sufficient to justify $\Delta \rho > \Delta\tilde{\rho}\ge 0$. To avoid this conclusion, the developer of a subquantum theory would have to solve a quite serious problem -- to obtain that the distortion $\delta\rho$, which is assumed to be maximal at $r=0$ by the principle of minimal distortion, nonetheless fulfills $\Delta\delta\rho=0$. This is clearly not what one would expect. 

\subsection{The subquantum energy balance}

But let's nonetheless consider another point -- what happens with the energy balance equation \eqref{balance}. Clearly, the quantum variant cannot be valid near the singularity. Subquantum theory has different equations, equations which have regular fields as solutions. There is no reason to modify the continuity equation, so that it is the energy balance equation which has to be modified.

And now let's enter some speculation about the cause of the regularization into the picture.  Subquantum theory has to allow non-zero $\w$, but to recover the quantum limit it needs a mechanism to suppress it, or, more accurate, to restrict it almost completely to the domain where density is low.  This may be reached, in particular, by a penalty term $U(\rho,\w)$ for non-zero $\w$. This may be something like $\rho(\w)^2$, but a particular choice of such a penalty potential is already beyond the scope of our purely qualitative considerations.

Nonetheless, some qualitative properties of such a penalty term seem almost obligatory: $U(\rho,\w)$ has to be positive for $(\w)^2>0, \rho>0$, and has to increase with $(\w)^2$. Now, if such a penalty term is incorporated into the balance equation, we obtain
\begin{equation}\label{modified}
\frac12 \tilde{\rho} \tilde{v}^2 + \frac12 \tilde{\rho} \tilde{u}^2 + U(\tilde{\rho},\nabla\times\tilde{v}) = \frac14\Delta \tilde{\rho}.
\end{equation}
Now, at $r=0$ the velocity-dependent terms disappear, but the penalty potential $U(\tilde{\rho},\nabla\times\tilde{v})$ does not: We have non-zero density $\tilde{\rho}(0)$, and $\nabla\times\tilde{v}$ is finite, and not only non-zero at $r=0$, but even maximal there. As a consequence, the value of $U(\tilde{\rho},\nabla\times\tilde{v})$ has to be positive. On the right hand side, we have only $\Delta \tilde{\rho}$, which, therefore, should be positive too. And, as we have already found, it is not greater than the original $\Delta\rho$, which, therefore, should be positive too.  

Thus, a purely qualitative consideration of the energy balance gives additional support for the hypothesis that $\Delta\tilde{\rho}> 0$, and, therefore, that $\Delta\rho>0$. 

\section{Discussion} 

The most important result is that to obtain empirical equivalence of quantum theory and the formulation of the theory in the flow variables $\rho(q),v^i(q)$ it is sufficient to add a simple regularity postulate for $\Delta\rho$ at the zeros of the density: All we need is $0<\Delta\rho<\infty$. 

The justification of the postulate remains, of course, speculative in character, because it is based on assumptions about an unknown subquantum theory. 

But these assumptions are consequences of a principle of minimal distortion, which is, in its methodological justification, similar to the rule of preference for the null hypothesis in statistics. So, on the general level, without specification of a particular subquantum theory, these assumptions do not require further justification in the same sense as independence assumptions do not require them. 

What remains is an additional consistency check for subquantum theories: In their quantum limit, not only the \Sch equation has to be derived, but also the regularity postulate \ref{p1}. Given the results of this paper, this will not be a serious problem for such theories -- if the theory does not violate the principle of minimal distortion, everything is fine. But, nonetheless, this remains to be checked if one wants to propose a subquantum theory. 

Given a particular subquantum theory, there would be also other related questions for future research: One would like to have a better understanding  what happens if we simply enforce non-integer values of $\a$ by the boundary conditions. The likely answer is that an excess curl will be much less localized than the quantized part of the curl, so that it may be distributed to other parts of the universe. Similarly, what happens with multiple zeros in a subquantum theory would be interesting too. 

But these interesting questions about particular subquantum theories do not change the main result of this paper: The Wallstrom objection is no longer a decisive objection against interpretations of quantum theories (as well as subquantum theories deriving them) in terms of flow variables. 

\begin{appendix}

\end{appendix}

\end{document}